# Codeathon Activity: A Design Prototype for Real World Problems

Preetha S[1], Rajeshwari K[1], Anitha C[1], Kausthub Narayan[1]

{preetha.ise@bmsce.ac.in, rajeshwarik.ise@bmsce.ac.in, anitha.bmsce@gmail.com, kausthubnaarayan@gmail.com}

**Abstract.** Activity-based learning helps students to learn through participation. A virtual codeathon activity, as part of this learning scheme, was conducted for 180 undergraduate students to focus on analysis and design of solutions to crucial real-world problems in the existing Covid-19 pandemic situation. In this paper, an analysis is made to know the problem solving skills of students given a single problem statement. Evaluators can further collate these multiple solutions into one optimal solution. This Codeathon activity impacts their practical approach towards the analysis and design.

*Keywords:* Software Engineering; Design Models; Codeathon; Education; Analysis; Collaboration; Activity-based Learning; Problem-based Learning; Alternate Assessment Tool; Multiple Solutions

## 1 Introduction

In this paper, an analysis is made on a codeathon activity conducted by the department of ISE, B.M.S. College of Engineering (BMSCE). The activity was conducted virtual for the theoretical course Software Engineering and Object Oriented Modelling (SEO). Software Engineering course discusses the various approaches used in solving real world problems to build a software product or a service. Object Oriented Modeling emphasizes on the design and modeling of the solutions for software programmers. As an autonomous institute, BMSCE provides faculties the liberty to frame the curriculum as well as to introduce new Teaching Learning Processes (TLP) for their courses. TLP is a well-planned method of establishing the learning objectives by creating implementation plans to meet the outcomes of the course. As part of TLP, Codeathon was planned as an Alternate Assessment Tool (AAT) as per evaluation scheme. Codeathon was conducted to motivate students, and to further explore creativity of students in problem solving techniques. Students learn better when they indulge in Activity-based Learning rather than reading from textbooks. This is a practical approach, where textbooks are used minimally and learning happens through activities and problem solving. Through Activity-based Learning (ABL) and Project-based Learning (PBL), students develop the skills to analyse and art of creativity gets nurtured. PBL is a method where students learn through projects, acquire knowledge and develop skills by working together in a team. Codeathon activity has been conducted for this course for four years in succession. The activity emphasized various real world problems whose solutions can be proposed in a day.

The students considered for analysis were from fifth semester, a total of 180 students (3 sections of 60 students) participated in this activity. Faculties composed teams of 3 or 4 students in a group; each section had around 15 batches. On a whole, the group consisted of a top performer, two average performers and a weak performer. The main intention was to have a different group, majorly a group without friends. This group could be diverse and can have compatibility issues initially to know each other. To know the strengths and weaknesses of the teammates, it would deliberately take more time. Also the acceptability of the solutions or strategies is more difficult when compared to a self-formed team. The course teachers were interested to learn about team-heterogeneity to encourage blended learning.

Due to pandemic, COVID-19, many challenges arose in the field of MSME [Micro, Small and Medium Enterprises], tourism and hospitality, retails, aviation, automobile, real estate sector, education and agriculture. The evaluators chose the most hit sector amongst these. The first one is the current system to which they belong to, i.e. education, others being the travel and tourism sector, and the prime part of daily consumption being agriculture. Each team had to analyse and model the solutions for these challenging problems. A scope was given to check how varied solutions could be obtained for a single problem. An attempt to solve a single problem by multiple teams led to the different solutions which converged to a single best solution. Students were given three main domains as problem statements, with two sub problems in each. Each section got one domain problem statement. The teams had to understand, analyse the problem and work together to find unique design solutions. For the evaluators, the submissions were made in stages, to know the progress on each solution and design techniques. Each team had to submit manuscripts describing the design and solution models they created during each stage.

The objective of this research is to understand how codeathon activity works and how their concept can be transformed to prototype solutions rather than software development. The key research goals are as follows:
- Development of a holistic understanding of codeathon activity
- Overview of AAT for course evaluation, setup of a draft concept for an activity-based learning, its conduction and analysis
- The synthesis of the results to improve TLP, its creativity, analysis, design thinking and modelling solutions for real world problems

For the last two years, the codeathon activity was conducted in the laboratory. Due to pandemic COVID-19, it was conducted virtually in this Academic Year [AY] [2020-2021]. The challenge of being distantly connected did affect the submission time and their performance. Codeathon is a best-suited model to synthesize and evaluate students, an ideal platform to learn collaboratively, and work to analyse the problem to showcase participants' skill sets. The activity's goal was concerned with design and blueprints, rather than coding or implementation of the solution. This practical approach showcases the uniqueness of idealistic solutions and how students adhere to the timelines. The teamwork process emphasized on the inclusivity and diversity amongst the participants with varied proficiency and domain knowledge.

## 2  Existing Literature

In [1], the study focused on investigating flow experiences of social interactions among small groups for positively sharing their experience. Individuals focus on their performance towards the activity and enjoy it during the activity. Collective flow can be observed when a group exhibits its highest capacities. A deep study on practices adopted towards conducting hackathons [2]. Hackathon events potentially serve a variety of purposes and identify the literature gaps related to hackathons. Significance of Hackathon was discussed in [3], study adopts historical and ethnographic techniques to determine work practices embraced by prestigious global industries. In hackathons, participants visualize themselves as representatives of social progress through software developments. Analysis and evaluation of design solutions using problem solving situations for a collaborative synchronous environment was presented in [4]. The Object-oriented Collaboration Analysis Framework (OCAF) model was proposed to analyse technology based collaborative problem solving methods. These methods are well suited for co-located group collaboration and synchronous distance-collaboration environments. Design solutions are developed using well distinguished objects, such as entity relationship diagrams, architectural diagrams, design formalisms, data flow diagrams and concept maps. Entities, relationships and attributes are considered as three basic constructs used to represent design solutions.

In [5] experience of teaching software engineering with a practical approach is discussed. A practical approach called "Software Engineering Project" (SEP) was conducted to understand project goals, domain requirements, technical background and team organisation. Honouring their responsibility is presented in [6]. The study inspires universities how software engineering apprehends completely into software curriculum which includes principles, practices, applications, tools and mathematics. In [7] General Social Survey (GSS) data is observed to figure out the relationship among racial diversity of workers' friendship networks and teamwork. Teams can nurture network formation and help to integrate through respect and cooperation. In [8], authors have examined how teaching practices, circumstances and a pool of initiatives correspond to enhancement of gender diversity in undergraduate software engineering programs. Department faculty were recognised to be the key to create diverse, inclusive and mentor social events /gatherings among students, also creating a platform among students for enhancing their skills. One of the most similar studies to our study is [9], the study observes the application of novel and interdisciplinary collaboration techniques which arose from the swiftly developing area of information technology. Also the hackathon is a pedagogical approach for collaboration. In [10], the study recommends impulsive communication as evidence for having close proximity in distributed workgroups and suggests structured management is suggested when groups lack cohesion. Various theoretical and methodological efforts are considered [11] to analyse the strengths of individuals, and a novel approach is proposed to address "diversity of diversity studies". The study in [12] observed the inclusion of diversity is displayed in courses offered across college curriculum. The authors depict how several course instructors adopt diversity in the course they teach through a comprehensive learning approach.

## 3 Encouraging Make-in-India Movement

Codeathon's theme was a miniature approach to the Make-in-India movement launched in 2014, by our Hon. PM Shri. Narendra Modi and the Ministry of Commerce and Industry. The government of India initiated this ambitious campaign to encourage development of solutions to manufacture in India. The campaign focuses on enhancement of skills and creation of jobs in 25 sectors as mentioned in [13]. The problem statements catered the initiations for this strategy. For the real world problems, the solutions should provide a framework and design the model for the implementation on a global platform. The solutions need to be feasible, innovative, user-friendly, to be modelled in a short span of time. Design models can further be accepted by the industrialists and thus drive the economy of our country.

A brainstorming session led to freezing of the problem statements. Various domains with greater challenges were considered. In Farming, how agriculture education can be improvised using AI, or to find people around specific places to know the total count who follow the Standard Operating Procedure (SOP) regulated by Government of India, for COVID-19 pandemic using voice recognition. Education during the COVID-19 was a greatest challenge, as it addressed a larger group of population across the world. Students needed Internet connectivity, basic infrastructure for the e-classes, and the assessments of examinations on the digital platform were more demanding for the course instructors. Aviation, travel were the next sectors to be considered. Due to the lockdown, inter-city travels became a nightmare. Keeping the front line workers safe was also a challenge.

## 4 Problem statement definitions

Three domains were chosen specifically to address the latest COVID-19 pandemic issues. These domains have experienced a catastrophic impact and have been the worst hit. A varied and feasible solution was expected from each section for the same. Students had the opportunity to further narrow down the domain and chose a sub-problem under it. The domains and the sub-problems were:

1. *Education Domain*
   - Conduction of Semester End Examinations (SEE) of Theory and practical sessions (On Campus examinations)
   - Theory Classes and Lab experiments conduction. Classes to be held and how college should maintain norms with SOP of Karnataka Government during COVID-19 pandemic
2. *Travel/Tourism Domain*
   - How to enforce Government's SOP during tourism and stay
   - Travel (any mode of transportation) during COVID-19 pandemic
3. *Agriculture Domain*
   - How technology can help farmers to bring Farm-to-Table
   - Enabling Seed-to-Table revolution

### 4.1 Stages of codeathon submissions

*Stage 1* expected the students to analyse the problem statement, to synthesize the Functional Requirements (FR) and Non-Functional Requirements (NFR)

*Stage 2* was a StarUML submission with high level diagrams in the form of Activity diagrams, Sequence diagrams, State chart diagrams or any Data flow diagrams. Also a use-case diagram if any. StarUML is a sophisticated software modeller for agile processes and concise modeling for diagramming. As the course was covering the design and modeling aspects, code implementations were not expected, where time was a concern

*Stage 3* helped teams to further justify their understanding of the problem by providing suitable test-suits which included failure analysis, verification and validation of the requirements

The whole process was a deeper dive into a higher level of modeling than a detailed implementation of the solution using any technology. Students had to nominate themselves into these roles: a Business analyst, Architect and a Quality & Assurance (Q&A). Each team was also provided with three flags to take expert advice from the evaluators. In table 1, a description of the stages and the submissions is briefed. As part of the assessment, evaluation timelines were informed according to the table description.

Table 1: Detailing of the submission stages

| Stages | No. of Eligible teams | Expected outcome (course concepts) | Mode of submission | Students Representation | Evaluators role/ flag usage |
|---|---|---|---|---|---|
| 1 | 45 | *Analysis phase:* decipher the problem statement requirements elicitation normal scenario, exceptional scenarios | PPT | Business analyst | Resolve ambiguities Team Assessment and promote for next stage or drop the team |
| 2 | 38 | *High Level Diagramming:* flow of the activities, modeling diagrams | StarUML submissions | Architect | Team Assessment, promote for next stage or drop the team |
| 3 | 26 | *Verification and validation:* Test Driven Development, failure analysis | MS Excel | Q&A | Choose best solutions to collate into one feasible solution for a problem statement |

Table 2: Rubric for evaluations

| Rubrics | Outcomes Expected | Stage Submissions |
|---|---|---|
| Problem Analysis | Understanding business models, analysing all requirements of the model to be developed. Elicit the requirements to develop unique solutions | Stage 1 |
| Analysis & Synthesis of FR and NFR | Able to analyse and synthesize FR and NFR for the system, in purview of the practical/feasible approach | Stage 1 |
| Designing High Level Diagram | Evaluate the requirements document artifact and come up with sequence diagrams and identifying use cases of the business/system | Stage 2 |
| Testing | Test cases are thorough and systematic, includes failure analysis as system meets performance requirements | Stage 3 |
| Team Work | Create unique Design Artifacts and solutions in-time and thrive at presentation | All stages |

### 4.2 Progression in Stages – The Rubrics

Students were graded based on their presentation, communication and novelty in the approach. Participants had to know about modeling at a high level design technique, how systems break up into modules, data flows, interaction modeling rather than checking schemas for object oriented modeling. In testing, a check was made to know how the model reacts to different scenarios, unexpected behaviour and exceptional scenarios; a tabulation on failure analysis was required. Table 2 depicts the evaluation parameters for the codeathon activity. The teams were evaluated in each stage based on these rubrics.

### 4.3 Discussion

The AAT focused on solutions for real world problems, aiming at team performance and involvement for the first year of its conduction in the campus, Academic Years [AY] [2017-18] with the industry expert guiding through the stages. AY [2018-19, 2019-2020] depicted a vivid working style of the teams and the creativity was improved. In AY [2020-2021], the mode of codeathon conduction was virtually, due to the ongoing pandemic, COVID-19.

Table 3 indicates the performance measure of students of different AY. It was observed that students in the AY 2020-21 were able to solve the problem more efficiently when compared to previous years. On the whole, the AAT proves a suggestive method to learn a theoretical course in an implementation style.

In this AY, the novelty of AAT is given a single problem, how multiple solutions can be achieved. As an evaluator, it was challenging to identify the best feasible solution amongst that problem. Teachers emphasized on the critical thinking and modeling which led to path breaking solutions. Though the problem statements were given on the spot, students were able to foresee the solutions apart from zero knowledge of the domain, without compromising performance.

**Table 3: Evaluation of the rubrics for different Academic Years**

| Evaluation Parameters | AY 2017-2018 | AY 2018-2019 | AY 2019-2020 | AY 2020-2021 |
|---|---|---|---|---|
| Problem Analysis | Medium | High | Medium | High |
| Analysis & Synthesis of FR and NFR | High | High | High | High |
| Designing High Level Diagram | High | Medium | Medium | Medium |
| Testing | Medium | Medium | High | High |
| Team Work | High | High | High | High |

## 5 Conclusions

The codeathon activity was conducted to converge multiple solutions of chosen three domains, how to collate feasible solutions into one best solution for each problem. The analysis highlights how difficult it is to obtain one solution for a domain. Strangely it was noted that the teams were not discussing the solutions amongst other teams in their section unlike the in-house scenario. Virtual activity forced them to focus completely towards team connectivity and meeting the deadline rather than discussing with peers. Evaluators picked the best solutions and converged them to a working prototype. The AAT in the course contributed for the student's understanding of the course in practical approach and the performance improvement. No matter what the situation is, students have proved that logical thinking and analysis will always lead towards better solutions. Current Academic Year, in particular, has tested students' capabilities to a great extent. Even though virtually connected, their effort towards problem solving was not compromised. AAT will be one of the prominent techniques towards concept understanding and applying theory concepts in dynamic problem solving.